\documentclass[preprint,12pt]{elsarticle}

\usepackage{amssymb}
\usepackage{amsmath}
\usepackage[T1]{fontenc}
\usepackage{graphicx}
\usepackage{enumitem}
\usepackage{xurl}
\usepackage[export]{adjustbox}
\usepackage{float}
\usepackage{booktabs}
\usepackage{appendix}
\usepackage{tabularx}

\usepackage{hyperref} 
\usepackage{color}
\usepackage{threeparttable}

\usepackage{listings}
\usepackage{xcolor}
\usepackage{upquote} 

\usepackage[most]{tcolorbox}

\newtcolorbox{procedurebox}[2][]{
  enhanced,
  breakable,
  colback=white,
  colframe=black,
  fonttitle=\bfseries,
  title={Procedure~\thetcbcounter: #2},
  before skip=10pt,
  after skip=10pt,
  #1
}

\journal{Science of Computer Programming}

\begin{document}

\begin{frontmatter}

\title{Automatic Generation of Formal Specification and Verification Annotations Using LLMs and Test Oracles\tnoteref{t1}}
\tnotetext[t1]{Preprint. Intended for submission to Science of Computer Programming.}

%% Authors
\author[inst1,inst2]{João Pascoal Faria}
\author[inst1]{Emanuel Trigo}
\author[inst1]{Vinicius Honorato}
\author[inst1,inst3]{Rui Abreu}

%% Author affiliations
\affiliation[inst1]{organization={Faculty of Engineering, University of Porto},
            country={Portugal}}

\affiliation[inst2]{organization={INESC TEC},
            city={Porto},
            country={Portugal}}
            
\affiliation[inst3]{organization={INESC-ID},
            city={Lisbon},
            country={Portugal}}

\begin{abstract}
Recent verification tools aim to make formal verification more accessible to software engineers by automating most of the verification process. However, annotating conventional programs with the formal specification and verification constructs (preconditions, postconditions, loop invariants, auxiliary predicates and functions and proof helpers) required to prove their correctness still demands significant manual effort and expertise. 
This paper investigates how LLMs can automatically generate such annotations for programs written in Dafny, a verification-aware programming language, starting from conventional code accompanied by natural language specifications (in comments) and test code.
In experiments on 110 Dafny programs, a multimodel approach combining Claude Opus 4.5 and GPT-5.2 generated correct annotations for 98.2\% of the programs within at most 8 repair iterations, using verifier feedback.
A logistic regression analysis shows that proof-helper annotations contribute disproportionately to problem difficulty for current LLMs.
Assertions in the test cases served as static oracles to automatically validate the generated pre/postconditions. We also compare generated and manual solutions and present an extension 
for Visual Studio Code to incorporate automatic generation into the IDE, with encouraging usability feedback.
\end{abstract}

% %%Graphical abstract
% \begin{graphicalabstract}
% \includegraphics[scale=0.7, trim={1.5cm 7.1cm 12.5cm 1cm},clip] {graphical_abstract.pdf}
% \end{graphicalabstract}

%%Research highlights
% \begin{highlights}
% \item A multimodel LLM approach generates Dafny pre/postconditions, invariants, and proof helpers.
% \item Achieves 98.2\% verification success on TESTDAFNY110 using repair prompting.
% \item Test assertions act as effective oracles for validating generated specifications.
% \item Generated pre/postconditions match expert solutions in 96.4\% of cases.
% \item Integrated into VS Code via Dafny AI Assistant, improving success rate and task time.
% \end{highlights}

%% Keywords

\end{frontmatter}

\section{Introduction}
\label{sec:intro}

Since software development is a knowledge-intensive human activity, increasingly assisted by ``untrusted'' AI/ML systems \cite{ahrendt2022trico}, and modern systems are increasingly complex, errors are almost inevitable; therefore, several techniques must be applied for defect prevention, detection, and resolution.
Testing can reveal bugs but cannot prove their absence \cite{dijkstra1970notes}, which limits its adequacy for critical systems \cite{boehm2006some}.
Formal verification addresses this gap by proving program correctness, but its required expertise, effort, and scalability challenges have restricted widespread use. Improving the accessibility of formal methods is thus essential for broader adoption \cite{ONCD2024}.

Recent advances in SAT and SMT solvers \cite{vardi2016automated}, such as Z3 \cite{moura2008z3}, enabled a new generation of verification tools that significantly improve the accessibility of formal methods for software engineers by automating large parts of the verification process. Representative examples include Dafny \cite{leino2017accessible}, Frama-C \cite{cuoq2012frama}, and Why3 \cite{filliatre2013why3}. Among these, Dafny stands out by providing a full-fledged programming language tightly integrated with specification and verification constructs, a Z3-based verifier, and IDE support through Visual Studio Code, resulting in a cohesive development environment that has seen increasing adoption in both industry and academia \cite{leino2017accessible}.
Nevertheless, despite these advances, the practical accessibility of Dafny and similar tools remains limited \cite{faria2023case}. A major obstacle is the effort and expertise required to write adequate formal specification and verification annotations, including not only preconditions, postconditions, loop invariants, and auxiliary predicates and functions, but also additional lemmas and assertions that are often necessary as proof helpers to compensate for the inherent limitations of automated verifiers.

Recent progress in large language models (LLMs), pre-trained on massive and diverse corpora \cite{brown2020language}, opens new opportunities for alleviating the annotation burden in formal verification. 

LLMs have already been explored for several verification-related tasks, including loop invariant inference \cite{kamath2023finding,pascoal2025automatic}, generation of verified Dafny methods from problem descriptions \cite{misu2024towards}, and suggestion of auxiliary lemmas \cite{silva2024leveraging}.

Building on these insights, our work leverages contemporary LLMs to
automatically generate all specification and verification annotations
required to prove the correctness of imperative programs written in Dafny, starting from conventional code with natural-language specifications and test cases with statically checked assertions, but no formal annotations.

\begin{figure}[H]
\begin{center}
\includegraphics[scale=0.6, trim={1.5cm 11.4cm 12.5cm 2.5cm},clip] 
{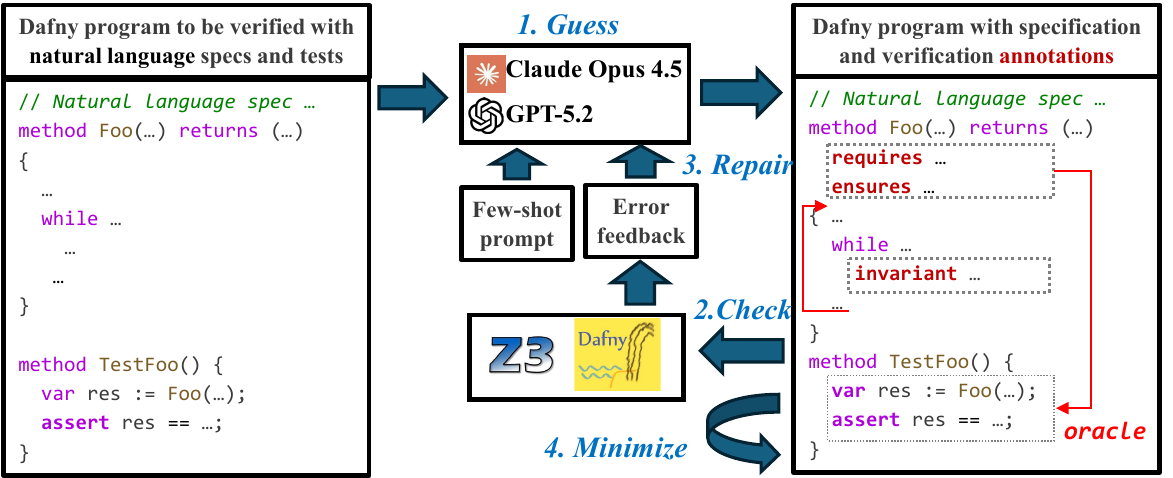}
\end{center}
\vspace{-12pt}
\caption{Proposed \textit{generate–check–repair-minimize} workflow for the automatic generation of specification and verification annotations using LLMs and test oracles. } 
\label{fig:approach}
\end{figure}

Unlike many LLM applications, our approach benefits from strong, machine checkable oracles, namely the Dafny SMT-based verifier and statically checked test assertions. These enable an iterative \textit{generate–check–repair-minimize} workflow in which LLM-generated annotations are automatically validated, refined using verifier feedback,
and subsequently minimized to remove redundant annotations (Fig~\ref{fig:approach}. This mitigates concerns about hallucination and trustworthiness, as only artifacts that pass formal verification are accepted.
We argue that this use case is particularly promising due to the availability of such complementary validation oracles, which naturally support the proposed workflow. Moreover, focusing on the generation of specification and verification annotations is more tractable than synthesizing complete
verified implementations \cite{misu2024towards, silva2024leveraging}, while the use of a high-level, verification-aware language such as Dafny enables addressing a broader class of problems compared to lower-level verification
settings \cite{kamath2023finding}.

The main contributions of this paper are:
\begin{itemize}[topsep=1pt,itemsep=1pt]
\item a curated dataset of 110 Dafny programs with accompanying test cases with statically checked test assertions  (TESTDAFNY110), publicly available on \href{https://github.com/joaopascoalfariafeup/testdafny110}{GitHub} (Sec.~\ref{sec:dataset});
\item an LLM-based approach for automatically generating specification and verification annotations for programs in Dafny, using test cases as specification oracles and carefully designed prompts and processing (Sec.~\ref{sec:prompt});
\item experimental results with contemporary LLMs, showing the effectiveness of our approach and the remaining challenges (Sec.~\ref{sec:exp} and Sec.~\ref{sec:results_and_discussion}), with details available in a  \href{https://github.com/joaopascoalfariafeup/testdafny110}{replication package};
\item Dafny AI Assistant, an extension to the Dafny plugin for Visual Studio Code that integrates specification and invariant generation in the IDE, publicly available on \href{https://github.com/emantrigo/dafny-plugin}{GitHub}, with promising user feedback (Sec.~\ref{sec:vscode}).
\end{itemize}

This paper is a significant extension of our previous work~\cite{pascoal2025automatic}, which focused only on generating loop invariants.
%(given pre/postconditions and proof helpers).

The remainder of the paper is organized as follows. 
Section~\ref{sec:inv} motivates the problem with an illustrative example. 
Sections~\ref{sec:dataset} (dataset), \ref{sec:prompt} (prompt engineering), 
\ref{sec:exp} (experimental design), \ref{sec:results_and_discussion} (results and discussion), 
and \ref{sec:vscode} (Dafny AI Assistant) present the main contributions. 
Related work is discussed in Section~\ref{sec:related}, and 
Section~\ref{sec:conclusions} concludes the paper.

\section{Program Verification in Dafny}
\label{sec:inv}

Dafny\footnote{\url{https://github.com/dafny-lang/dafny}}\cite{leino2017accessible, leino2023program} is a multi-paradigm programming language for creating verified programs based on design-by-contract. It uses a functional style for specifications (pure functions, predicates, and expressions) and procedural and object-oriented styles for implementations (methods and statements with side effects). Dafny is supported by a Z3-based verifier, compilers for C\#, Java, JavaScript, Go, and C++, and a Visual Studio Code extension.

\begin{figure}
\begin{center}
\includegraphics[scale=0.78, trim={3cm 1cm 13.2cm 0.5cm},clip] {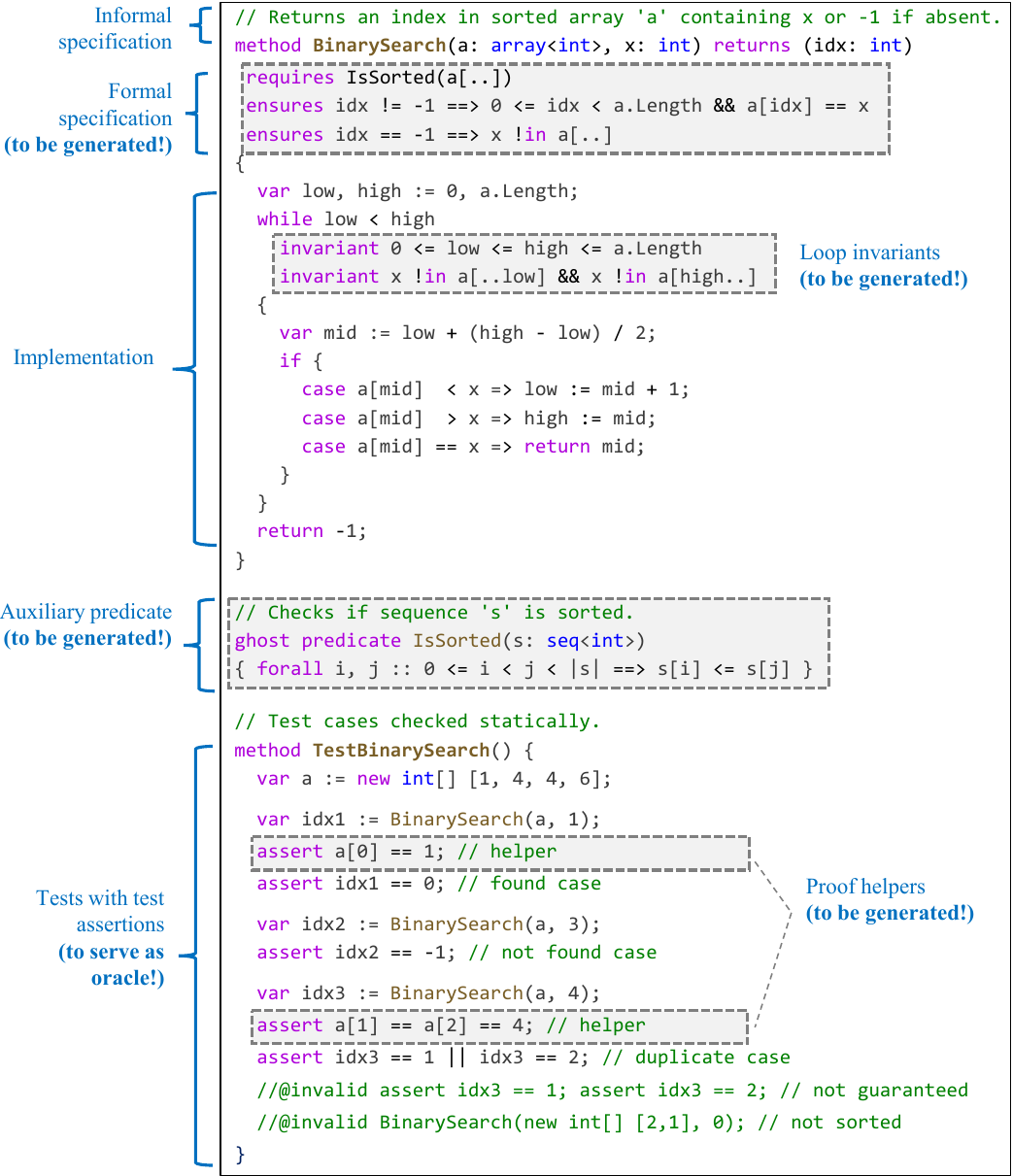}
\end{center}
\vspace{-8pt}
\caption{Example of a program in Dafny with formal annotations required for successful verification by the Dafny verifier. Our goal is to generate automatically such annotations.} 
\label{fig:div}
\end{figure}

Fig. \ref{fig:div} shows an example of a program in Dafny with a core method under verification (\texttt{BinarySearch}), a test method (\texttt{TestBinarySearch}) with relevant test cases and test assertions, and several specification and verification annotations required for proving program correctness with the Dafny verifier: 
\begin{itemize}[topsep=2pt,itemsep=1pt]
\item formal specification of the method's \textit{preconditions} and \textit{postconditions}, with \texttt{requires} and \texttt{ensures} clauses;
\item an auxiliary \textit{ghost predicate} (\texttt{IsSorted}) used in the precondition;
%~\footnote{Auxiliary \textit{ghost functions} are also commonly used.}
\item \textit{loop invariants} (with the \texttt{invariant} clause), to help the verifier check the method implementation against the specification; \footnote{In some cases, we may also need to provide a \textit{loop variant} with the \texttt{decreases} clause.}

\item \textit{proof helpers}, in the form of auxiliary assertions (with \texttt {assert}), to help the Dafny verifier accomplish its job.\footnote{Other types of more advanced proof helpers are \textit{lemmas} and \textit{ghost variables}.}   

\end{itemize}

A loop invariant (or inductive loop invariant \cite{kamath2023finding}) is a Boolean condition that holds at the loop entry
and after each iteration, and is strong enough to imply the method postcondition when the loop terminates. 

Our goal is precisely to relieve the user from providing such specification and verification annotations whenever possible, by leveraging the capabilities of LLMs, further automating program verification.
\footnote{Developers may first create 'traditional' programs using \texttt{\{:verify false\}} and \texttt{expect} (checked at runtime) instead of \texttt{assert}, and subsequently proceed to formal verification.}

In our dataset, each method is accompanied by test cases with \texttt{assert} statements, which the Dafny verifier checks statically using only the method’s pre- and postconditions and ignoring the body. These \emph{static test cases} therefore act as specification-level oracles, validating specifications at compile time rather than implementations at run time, and are effective at revealing incomplete postconditions. For example, a postcondition such as \texttt{0 <= idx < a.Length ==> a[idx] == x} would fail to verify the test assertions, since values outside \texttt{-1 <= idx < a.Length} would also satisfy it.

Due to limitations of automated theorem provers, Dafny verification often requires proof helpers. In test code, a common helper (Fig. \ref{fig:div}) is the use of auxiliary assertions to justify the test assertions.

\section{Curated Dataset}
\label{sec:dataset}

To assess the effectiveness of LLM-based generation of formal specification and verification annotations, we created a composite dataset with 110 programs in Dafny, named TESTDAFNY110, combining three subdatasets:
\begin{itemize}[topsep=1pt, itemsep=1pt]
\item{Subdataset A - refined versions of 85 programs from the \href {https://github.com/Mondego/dafny-synthesis}{MBPP-DFY-153 dataset} \cite{misu2024towards}, with solutions in Dafny to programming problems defined in the widely-used MBPP dataset \cite{austin2021program}, which is a standard benchmark for evaluating the code generation capabilities of LLMs;} 
\item{Subdataset B - 15 programs from the \href{https://github.com/joaopascoalfariafeup/loopinv100}{LOOPINV100 dataset} described in our previous work \cite{pascoal2025automatic}, including programs of varying complexity sourced from academic challenges (numerical algorithms, searching and sorting algorithms, etc.), to enhance the diversity of the dataset;}
\item{Subdataset C - 10 new programs of varying complexity, with the goal of reducing the likelihood of prior exposure during model training.
\footnote{The LOOPINV100 dataset in our previous work \cite{pascoal2025automatic} included subsets A and B; this work adds subset C with 10\% new problems.}}

\end{itemize}

As explained in our previous work \cite{pascoal2025automatic}, we selected and significantly refined with static test assertions all loop-based programs from the MBPP-DFY-153 dataset, excluding duplicates and cases that could be solved trivially without loops, resulting in 85 programs. 

A distinctive characteristic of the TESTDAFNY110 dataset is that all programs include test cases with statically checked \texttt{assert} statements, so that they can serve as test oracles to uncover issues with the specification.

The TESTDAFNY110 dataset is available in a public  \href{https://github.com/joaopascoalfariafeup/testdafny110}{GitHub repository}\footnote{\url{https://github.com/joaopascoalfariafeup/testdafny110}}.
Table \ref{tab:optypes} summarizes some of its characteristics. 

\begin{table}[H]
\caption{Composition of the TESTDAFNY dataset, with the number of programs (P), lines of code (LOC) excluding blank lines, comments and annotations (L), LOC representing annotations (A), and annotations representing proof helpers (H) per problem category.}
\label{tab:optypes}
\footnotesize
\begin{tabular}{|c|p{7.2cm}|c|c|c|c|}
\hline
\textbf{Categ.} & \textbf{Description} & \textbf{P} & \textbf{L} & \textbf{A} & \textbf{H}\\
\hline
check &Check if a scalar (\texttt{int}, \texttt{nat}, \texttt{real}) or list (\texttt{array}, \texttt{seq} or \texttt{string}) satisfies a condition (\texttt{IsPrime}, \texttt{IsMinHeap}, \texttt{IsPerfectSquare}, etc.).&22 & 507& 88 &33\\ \hline
%\texttt{IsSubList}, 
filter &Retrieve a sublist or subset according to a predicate (\texttt{RemoveDuplicates}, \texttt{Intersection}, \texttt{Difference}, \texttt{FilterOddNumbers}, etc.).& 11& 274 & 142 & 37\\ 
%\texttt{FindNegativeNumbers}, 
\hline
map &Map lists, element-wise or pairwise, into a new list of equal length (\texttt{ToUppercase}, \texttt{ElementWiseAddition},  etc.).&21& 404& 114 & 4 \\ \hline
math &A mathematical operation or recurrence (\texttt{Fibonnacci},   \texttt{FastModularExponentiation}, \texttt{CombNK}, \texttt{PrimeFactorization}, etc.). & 9 & 240 & 316 & 16 \\ \hline
merge &Combine two or more lists or a list and other data into a new list (\texttt{Merge}, \texttt{Interleave}, \texttt{IntersectIntervals}, etc.).&5& 96& 53 & 11\\ \hline
reduce &Retrieve an aggregate value from a list, sublist, or range  (\texttt{CountTrue}, \texttt{SumRange}, \texttt{SumNegatives}, \texttt{MaxDifference}, \texttt{Mode}, etc.).&15& 341& 175 &29 \\ \hline
reorder &Sort, reorder or reorganize the elements in a list (\texttt{BubbleSort}, \texttt{InsertionSort}, \texttt{RotateRight}, \texttt{PartitionOddEven}, etc.). & 10 & 229 & 158  & 20 \\ \hline
search &Obtain a value or location in a list based on a search criterion (\texttt{BinarySearch}, \texttt{FindFirstOccurrence}, \texttt{FindMax},  etc.).&17& 414& 165 & 38\\ \hline
\textbf{Total}&& \textbf{110} & \textbf{2505} & \textbf{1211} & \textbf{188}\\
\hline
\end{tabular}
% * The \texttt{FastModularExponentiation} and \texttt{PrimeFactorization} are the most challenging problems in our dataset, requiring 4 and 6 lemmas, respectively, and a high number of annotations in the manual solution.
\end{table}

\section{Prompt Engineering and Result Processing}
\label{sec:prompt}

We iteratively designed two system prompts:
\vspace{-6 pt} 
\begin{itemize}
\item{\textbf{Direct prompt} (\ref{sec:base_prompt}) - Instructs the LLM to generate the annotations required for successful verification of a Dafny program. }
\item{\textbf{Repair prompt} (\ref{sec:repair_prompt}) — Instructs the LLM to revise existing annotations in a Dafny program to achieve successful verification, using the error information reported by the Dafny verifier.}
\end{itemize}

\vspace{-3 pt} 

Most programs in our dataset (56\%) require proof helpers (assertions, lemmas) for successful verification beyond pre/postconditions and loop invariants, especially in test code, but their need is difficult to predict upfront due to the verifier’s heuristics~\cite{faria2023case}. In practice, developers first write the required pre/postconditions and loop invariants and then iteratively add proof helpers based on the verifier’s error feedback. Our two prompts are designed to mirror this workflow in an LLM-assisted setting.

To design the direct prompt, we started from a minimal zero-shot prompt and iteratively refined it by adding instructions to prevent syntax and semantic errors and guide the model in the derivation of pre/postconditions and loop invariants, including generic code patterns (few-shot prompting).

\begin{figure}[H]
\begin{center}
\includegraphics[scale=0.67, trim={1.2cm 14.1cm 13.5cm 0.2cm},clip, center] {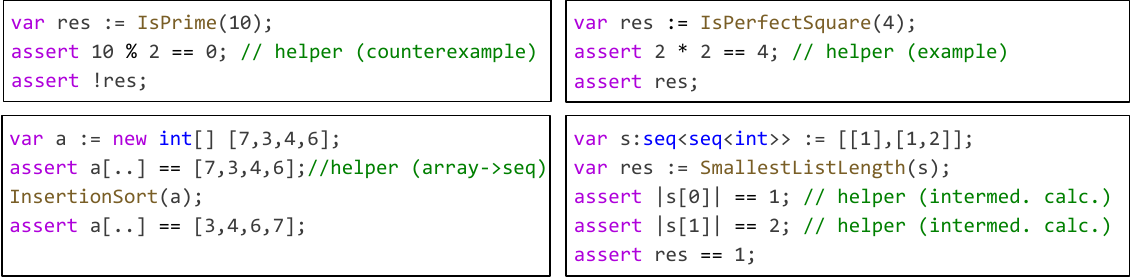}
\end{center}
\vspace{-12pt}
\caption{Examples of common patterns of proof helpers needed in test code.} 
\label{fig:testhelpers}
\end{figure}

We designed the repair prompt based on the observation that many proof helpers follow recurring patterns, such as assertions over basic sequence properties (see part~3a of the repair prompt) and auxiliary assertions in test methods (see Fig.~\ref{fig:testhelpers}). Accordingly, the repair prompt conveys an evolving catalogue of common repair strategies. We also introduced guardrails to prevent the LLM from \emph{cheating}, by inserting unproved \texttt{assume} statements, disabling termination checking (with \texttt{decreases *}), or removing test assertions.

We also implemented post-processing steps to address recurring issues in the LLM-generated outputs. For example, despite explicit instructions, loop invariants are sometimes placed incorrectly, and need to be relocated.

Especially under iterative repair prompting, LLM-generated solutions tend to be significantly longer than manually written ones, due to the accumulation of patches (typically auxiliary proof steps or helper assertions) introduced across successive repair iterations.
To address this issue, we developed an automatic \textbf{minimization procedure} (Section~\ref{sec:minimization}) that can be applied to already verified solutions.
The procedure iteratively removes generated code segments, retaining each removal only if both verification success and verification time are preserved.

\section{Experimental Design}
\label{sec:exp}

\subsection{Research Questions}

We aim to answer the following research questions:
\begin{itemize}
\item \textbf{RQ1.} How effectively can current LLMs generate specification and verification annotations (pre/postconditions, loop invariants, and proof helpers) to successfully verify Dafny programs of moderate complexity?
\item \textbf{RQ2.} How effective are test methods with test assertions at detecting incorrect or incomplete specifications?
\item \textbf{RQ3.} How does the quality of LLM-generated solutions compare to expert solutions?
\item \textbf{RQ4.} What are the main challenges and limitations of LLM-guided generation of specification and verification annotations?
\end{itemize}

\subsection{Methods}

We conducted two types of experiments to assess the effectiveness of the direct prompt (\textit{direct prompting strategy}) and the repair prompt (\textit{repair prompting strategy}).

In the direct prompting strategy, we removed all annotations (pre/postconditions, loop invariants, lemmas, ghost functions and predicates, and helper assertions) from the 110 programs in the TESTDAFNY110 dataset, retaining only the assertions in test methods that check test outcomes. Each stripped program (representing a conventional program) was then provided to the model via the corresponding API, with the goal of generating the specification and verification annotations required for successful verification by the Dafny verifier.
For each program, we performed up to $n=5$ independent runs to account for the stochastic nature of LLM outputs, stopping early if verification succeeded.

In the repair prompting strategy, the first attempt followed the direct prompting strategy. If verification failed, we performed up to nine repair iterations using the repair prompt, each time providing the current annotated program together with the error information reported by the verifier, and stopping upon successful verification.

We checked each generated output (after post-processing) for correctness using the Dafny verifier (version~4.11.0) with a timeout of one minute. Since the verifier also checks the assertions in the test code, this also means that the generated specifications (pre/postconditions) pass the tests. Later, we discuss the effectiveness of this criterion. 

We experimented with state-of-the-art models from different providers (Antrophic, OpenAI and DeepSeek), plus a model with an older knowledge cutoff date (for assessing its impact). The models, snapshots, release dates (R), and knowldege cuttof dates (C) are: 

\begin{itemize}
\item{Claude Opus 4.5 (claude-opus-4-5-20251101), R=11/2025, C=05/2025;}
\item{GPT-5.2 (gpt-5.2-2025-12-11), R=12/2025, C=10/2025 (reasoning model);}
\item{DeepSeek-V3.2 (chat), R=12/2025, C=07/2024 (lower cost model);}
\item{GPT-4 (gpt-4-0613), R=03/2023, C=09/2021 (older model).}
\end{itemize}

Except for GPT-5.2, which uses a fixed temperature of 1.0, all models were evaluated with a temperature of 0.5, following the findings in prior work~\cite{misu2024towards}. To assess the impact of this parameter, we additionally evaluated the top-performing model (Claude Opus~4.5) with a temperature of 0.

GPT-5.2 was tried at two reasoning effort levels - none and low - to analyze the effect of this parameter.

Given the complementary strengths of Claude Opus 4.5 and GPT-5.2 (the best performing models), we also tried a \textit{multimodel} parallel combination. In each attempt, we run both models and select the best output, prioritizing (in order) (i) syntactically valid outputs, (ii) that pass the verifier, (iii) have fewer LOC, and (iv) are verified in less time.

For each model configuration, we measured the percentage of problems solved within up to $k$ attempts. A problem is considered solved if the LLM output passes the Dafny verifier. We denote this metric as $pass@k$ under direct prompting and $repair@k$ under repair prompting.
We also record API response times and costs, given their practical relevance. For verified programs, we measure solution complexity by counting the number of lines of code, enabling comparison with manual solutions.

The experiments were automated by a Python script (available in our \href{https://github.com/joaopascoalfariafeup/testdafny110}{TESTDAFNY110 repository}) run in an Intel(R) Core(TM) Ultra 7 258V CPU @2.20 GHz, with 8 cores and 32 GB RAM running Windows 11 Pro.

\section{Experimental Results and Discussion}
\label{sec:results_and_discussion}

\subsection{Performance Results}

\begin{figure} [H]
\begin{center}
\includegraphics[scale=0.7, trim={2.1cm 9cm 11.9cm 2.6cm},clip] {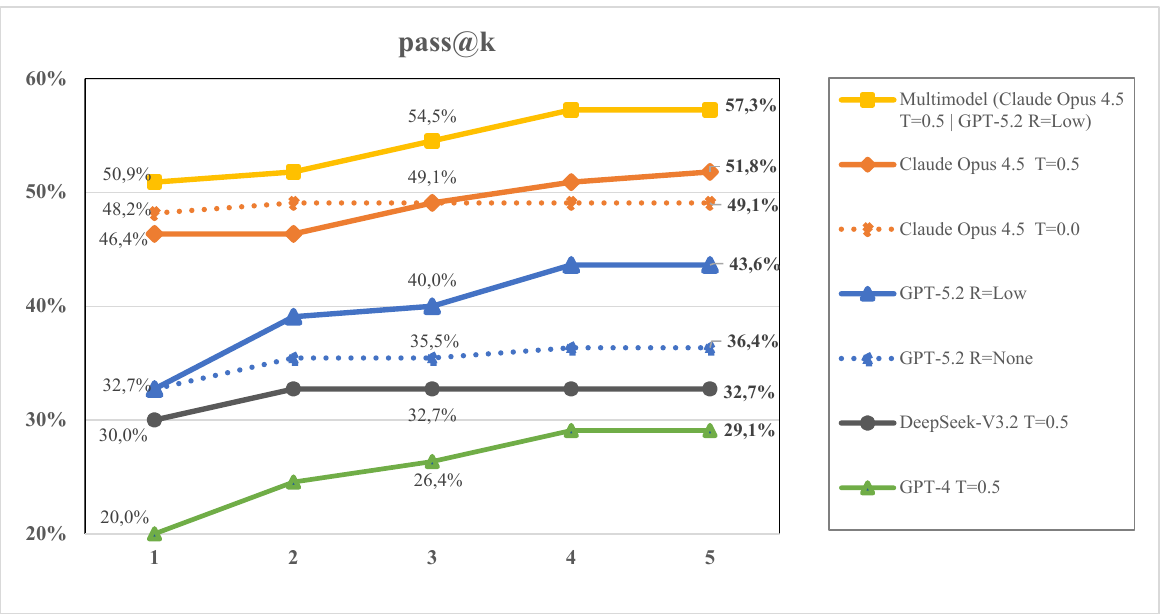}
\end{center}
\vspace{-6pt}
\caption{Percentage of problems of the TESTDAFNY110 dataset solved after $k$ attempts ($pass@k$) in the direct prompting strategy using different models and configurations.}
\label{fig:passk}
\end{figure}

\begin{figure} [H]
\begin{center}
\includegraphics[scale=0.7, trim={2.1cm 9cm 11.9cm 2.6cm},clip] {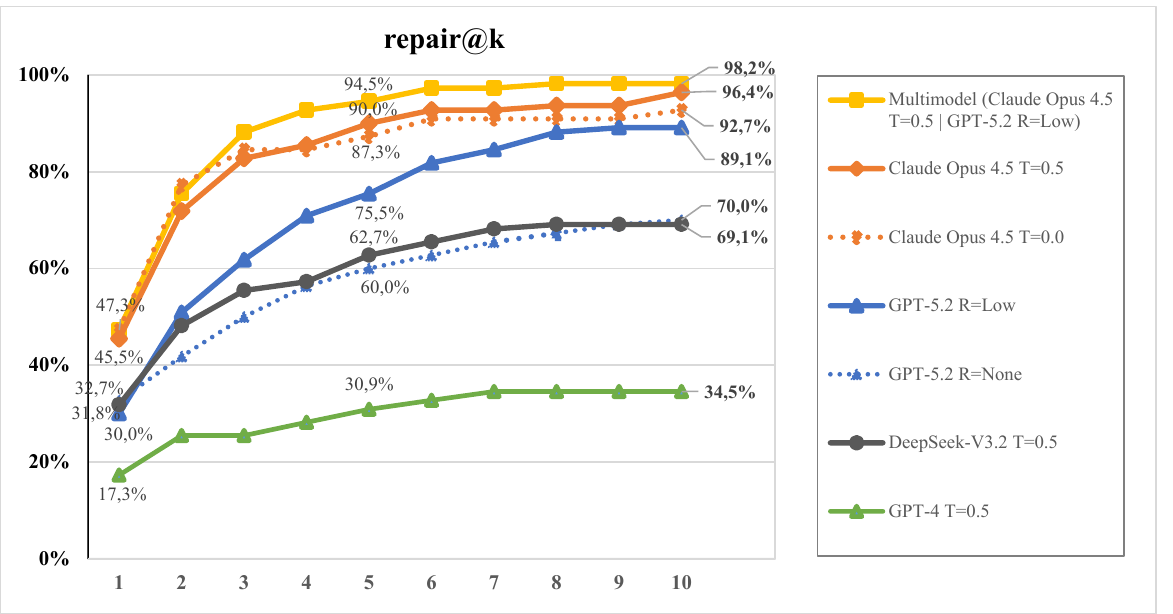}
\end{center}
\vspace{-6pt}
\caption{Percentage of problems of the TESTDAFNY110 dataset solved after $k$ attempts ($repair@k$) in the repair prompting strategy using different models and configurations.}
\label{fig:repairk}
\end{figure}

Figs. \ref{fig:passk} and \ref{fig:repairk} display the value of the $pass@k$ and $repair@k$ metrics for all the models and configurations tried, with the direct prompting and repair prompting strategy, respectively. Table \ref{tab:modelperf} summarizes the performance results for all the models, configurations, and prompting strategies. 
Raw results are available in our \href{https://github.com/joaopascoalfariafeup/testdafny110}{TESTDAFNY110 repository}.

\begin{table} [H]
\caption{Performance results for different models, configurations (T-temperature, R-reasoning effort) and prompting strategies, showing success rates (p@k=pass@k or r@k=repair@k), average cost and response time per API call, and percentage of extra LOC compared to the manual solution in successful attempts.}
\label{tab:modelperf}
\centering
\footnotesize
\begin{tabular}{|p{4.7cm}|p{0.9cm}|p{0.9cm}|p{0.9cm}|p{1.3cm}|p{0.9cm}|p{0.9cm}|}
\hline
\textbf{Model} & \textbf{p@1/ r@1} & \textbf{p@5/ r@5}  & \textbf{-/ r@10} & \textbf{Cost (\textcent~USD)} & \textbf{Time (s)} & \textbf{Extra LOC}\\
\hline
\textit{Direct prompting strategy:}& & & & & & \\
Multimodel (CO 4.5|GPT-5.2)& \textbf{50.9\%} & \textbf{57.3\%} & - & 4.55 & \textit{24.4} & 7.9\%\\
Claude Opus 4.5 T=0.5& 46.4\% & 51.8\%  & - & 2.28 & 8.2 & 8.8\%  \\
Claude Opus 4.5 T=0& 48.2\% & 49.1\%  & - & 3.21 & 8.7 & 11.7\%  \\
GPT-5.2 R=Low& 32.7\% & 43.6\% & - & 2.26 & 23.7 & \textit{14.0\%} \\
GPT-5.2 R=None& 32.7\% & 36.4\% & -  & 0.71 & \textbf{4.5} & 4.8\% \\
DeepSeek-V3.2 T=0.5& 30.0\% & 32.7\% & - & \textbf{0.03} & 12.9 & 5.1\% \\
GPT-4 T=0.5& \textit{20.0\%} & \textit{29.1\%} &  - & \textit{6.80}  & 12.0 & \textbf{1.2\%} \\
\hline
\textit{Repair prompting strategy:}& & & & & & \\
Multimodel (CO 4.5|GPT-5.2)& \textbf{47.3\%} & \textbf{94.5\%} & \textbf{98.2\%} & 7.86 &28.5 & 63.2\%*\\
Claude Opus 4.5 T=0.5& 45.5\% & 90.0\%  & 96.4\% & 4.41 & 15.8 & 50.4\%  \\
Claude Opus 4.5 T=0& 47.3\% & 87.3\%  & 92.7\% & 4.51 & 16.5 & 39.7\%  \\
GPT-5.2 R=Low& 30.0\% & 75.5\% & 89.1\% & 3.45 & 29.5 & \textit{153\%} \\
GPT-5.2 R=None& 32.7\% & 60.0\% & 70.0\%  & 3.51 & \textbf{9.1} & 29.4\% \\
DeepSeek-V3.2 T=0.5& 31.8\% & 62.7\% & 69.1\% & \textbf{0.07} & 31.5 & 25.0\% \\
GPT-4 T=0.5& \textit{17.3\%} & \textit{30.9\%} &  \textit{34.5\%} & \textit{8.80}  & 15.2 & \textbf{3.3\%} \\
\hline
\end{tabular}
* Reduced to 10.6\% by applying the minimization procedure. 
\end{table}

As expected (given that the majority of the programs might require proof helpers), success rates are much higher in the repair prompting strategy, as compared to the direct prompting strategy.

Individually, the best-performing model is Claude Opus 4.5, with very similar success rates across different \textbf{temperature (T)} settings, yielding better results with T=0.5 after a few iterations.

To assess the impact of the \textbf{reasoning effort level (R)}, we tested GPT-5.2 with R=Low and R=None, obtaining better success rates with R=Low after a few initial iterations, at the expense of longer response times and more verbose solutions (see Table~\ref{tab:modelperf}).

By combining Claude Opus~4.5 (with $T{=}0.5$) and GPT-5.2 (with $R{=}$Low) in parallel, \textbf{our multimodel approach with repair prompting solved 98.2\% of the problems within at most 8 attempts ($repair@8$), requiring only 2 attempts on average [RQ1]}. Only two complex problems were not solved: \texttt{FastModularExponentiation} and \texttt{PrimeFactorization}, requiring 4 and 10 lemmas in the manual solutions, respectively.

The higher success rates of repair prompting come at the cost of larger solutions: generated programs are 63.2\% larger than manual ones, compared to only 7.9\% with direct prompting. By applying the automatic \textbf{minimization} procedure (\ref{sec:minimization}) to verified solutions it \textbf{removed 58\% of the annotation LOCs}, and \textbf{reduced the LOC overhead over manual solutions from 63.2\% to 10.6\%}, taking 1.4~s per line removed on average.

Considering all performance dimensions shown in Table \ref{tab:modelperf}, we conclude that \textbf{Claude Opus 4.5 with T=0.5 provided the best balance across all relevant metrics}, with $repair@5=90.0\%$, $repair@10=96.4\%$, and an average cost of 4.41~\textcent{} and response time of 15.8~s per API call in the repair prompting strategy, making it adequate for practical use. By comparison, the Dafny verifier took on average 9.7~s per verification task (1.9~s when timeout cases are excluded).

Considering API cost or response time alone, \textbf{the model with the lowest cost was DeepSeek-V3.2} (0.07~\textcent{} per call on average in the repair prompting strategy), by a large margin, while the model with the lowest average response time (9.1~s per call in the repair prompting strategy), by a small margin, was GPT-5.2 with R=None (see Table~\ref{tab:modelperf}).

 To assess possible \textbf{data contamination}, we additionally evaluated GPT-4, whose knowledge cutoff predates the source repositories used in our dataset. GPT-4 achieves a $pass@5$ score that is 7.3 percentage points lower than that of OpenAI’s newest model (GPT-5.2) in non-reasoning mode (29.1\% vs. 36.4\% in Fig.~\ref{fig:passk}). This performance gap between model generations, which is consistent with observations on contamination-resistant benchmarks such as LiveBench~\cite{livebench}, suggests a low contamination risk in our evaluation.
Under the repair prompting strategy (Fig.~\ref{fig:passk}), the performance gap is substantially larger, likely reflecting the limited ability of older models to effectively exploit iterative error feedback.

\subsection{Distribution of Error Types Across LLMs}

\begin{figure} [H]
\begin{center}
\includegraphics[scale=0.92, trim={2.5cm 13.6cm 16.1cm 3.1cm},clip, center] {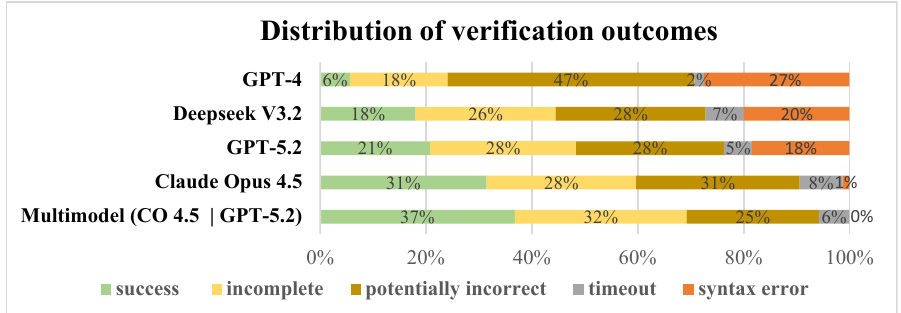}
\end{center}
\vspace{-12pt}
\caption{Distribution of verification outcomes across LLMs.} 
\label{fig:errortypes}
\end{figure}

Figure~\ref{fig:errortypes} shows the distribution of success rates and error types across LLMs, aggregated over prompting strategies and configurations (temperature and reasoning effort). Success rates appear low because all generation attempts are considered, including unsuccessful repair iterations.

Error types were classified automatically. \textit{Syntax errors} are identified when the verifier fails with resolution/type or parse errors. Otherwise, the model outputs are merged with the manual solutions to identify \textit{incomplete} solutions whenever verification succeeds.\footnote{Due to limitations of the verifier and the merge procedure, there may exist a higher percentage of incomplete outputs (weak pre/postconditions and loop invariants).} The remaining cases are classified as \textit{timeout} if the verifier times out, or \textit{potentially incorrect} if verification fails.

As expected, \textbf{older models produce more syntax errors (27\% with GPT-4), whereas more recent and capable models practically do not have syntax errors (1\% with Claude Opus 4.5) [RQ1]}.

\subsection{Analysis of Success Factors}
\label{sec:factors}

In this section, we analyze which factors influence success rates by fitting a simple and interpretable probabilistic model over structural properties of the programs, evaluated across the 14 model configurations listed in Table~\ref{tab:modelperf}.

We consider three structural features extracted from the manual solutions: \texttt{L}, the number of lines of code in the original program (excluding blank lines, comments, and annotations); \texttt{A}, the number of annotation lines; and \texttt{H}, the number of proof-helper annotations (23, 9, and 2, respectively, in Fig.~\ref{fig:div}).

Annotation generation success (defined as the ability to produce at least one output that passes the Dafny verifier within $k=5$ and $k=10$ attempts for direct and repair prompting, respectively) is modeled as a binary outcome using \textit{logistic regression}, with structural features as predictors and model configuration as a categorical factor.
For program $i$ and configuration $j$, the success probability is defined as:
\vspace{-6pt}
\[
\Pr(\text{success}_{i,j}) =
\frac{1}{1 + \exp\!\left(-\big(
\beta_0
+ \beta_L\,\texttt{L}_i
+ \beta_A\,\texttt{A}_i
+ \beta_H\,\texttt{H}_i
+ \alpha_j
\big)\right)},
\]
where $\beta_L$, $\beta_A$, and $\beta_H$ are shared coefficients capturing the influence of structural features, and $\alpha_j$ is a configuration-specific intercept.
This formulation yields a single, shared notion of problem difficulty across configurations, while allowing each model to differ only in baseline success probability.

\begin{figure}[H] 
\begin{center} 
\includegraphics[width=0.43\linewidth]{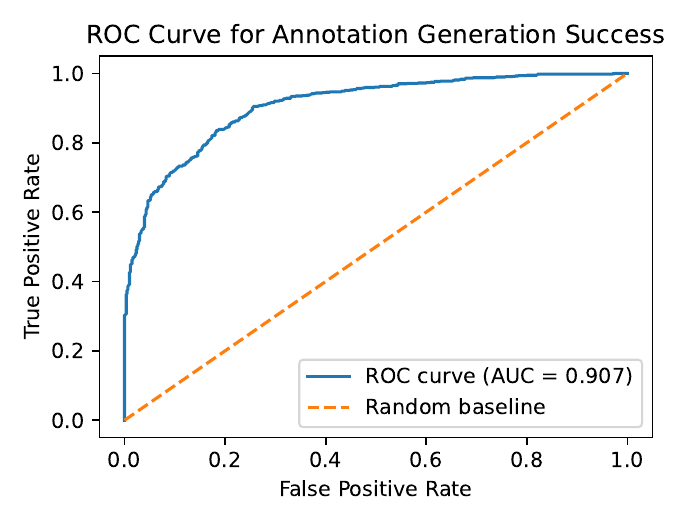} 
\end{center} 
\vspace{-24pt} 
\caption{ROC curve for the logistic regression model predicting LLM-generation success across all 1,540 problem--configuration pairs (110 programs $\times$ 
14 configurations).}
\label{fig:roc_curve} 
\end{figure}

Figure~\ref{fig:roc_curve} reports the receiver operating characteristic (ROC) curve aggregated over all problem--configuration pairs.
The resulting area under the curve (AUC=0.907) indicates strong discriminative performance. AUC values computed separately for each configuration are consistently high (0.80--1.00), confirming that the learned difficulty signal generalizes across models.

The learned coefficients show that success probability decreases with program size ($\beta_L=-0.0735$), annotation size ($\beta_A=-0.0475$), and most strongly with the number of proof-helper annotations ($\beta_H=-0.5825$); all effects are significant at $p<0.001$ (Wald tests). The larger magnitude of $\beta_H$ indicates that \textbf{proof-helper annotations contribute disproportionately to problem difficulty compared to standard specification elements for current LLMs [RQ4].}

To assess potential effects of prior exposure, we added categorical indicators for subdataset membership (Section~\ref{sec:dataset}). These factors neither improved model fit (likelihood-ratio test, $\chi^2(2)=0.79$, $p=0.67$) nor discriminative performance (AUC 0.907 vs.\ 0.908). 
%The coefficients were not statistically significant, indicating that program recency has no effect once structural complexity is controlled for.
The subdataset coefficients were not statistically significant, suggesting that program recency does not affect verification success once structural complexity is controlled for. 

Overall, these results show that \textbf{variation in success rates is primarily explained by structural complexity rather than by subdataset origin or recency}, supporting the robustness of the evaluation and mitigating concerns about data contamination or prior exposure.

\subsection{Quality of Generated Specifications and Role of Test Assertions}
\label{sec:specquality}

To assess the quality of the LLM-generated specifications, we manually compared the generated pre/postconditions (after minimization) for all 110 programs against the manual solutions, with the results shown in Table~\ref{tab:assessment}.

\vspace{-6pt}
\begin{table} [H]
\caption{Comparison of generated pre/postconditions to expert solutions.}
\label{tab:assessment}
\footnotesize
\centering
\small
\begin{tabular}{|p{2.8cm}|p{7.4cm}|p{2.3cm}|}
\hline
\textbf{Category} & \textbf{Description} & \textbf{\#Programs} \\
\hline
\multicolumn{2}{|l|}{\textbf{Logically equivalent}} & \textbf{106 (96.4\%)}\\
\hline
Same wording & Textually identical expressions.& 12 (10.9\%)\\
\hline
Same syntax & Structurally identical (different names). & 11 (10.0\%) \\
\hline
Same semantics & Different expressions, similar complexity. & 68 (61.8\%) \\
\hline
More complex & More complex than the manual solution. & 8 (7.3\%) \\
\hline
Less complex & Simpler than the manual solution. & 2 (1.8\%) \\
\hline
Less declarative & Closer to the implementation (see Fig. \ref{fig:countdistinct}). & 3 (2.7\%) \\
\hline
More declarative & Closer to natural language specification. & 2 (1.8\%) \\
\hline
\multicolumn{2}{|l|}{\textbf{Not logically equivalent}} & \textbf{4 (3.6\%)}\\
\hline
Stronger postconditions & Stronger than specified in NL,  but consistent with implementation (Fig.~\ref{fig:linearsearch}). & 1 (0.9\%) \\
\hline
Weaker precondition & Waker than specified in NL, but consistent with the implementation. & 3 (2.7\%) \\
\hline
 % \textbf{Total} & & \textbf{110} \\
 % \hline
\end{tabular} 
\end{table}

\vspace{-6pt}

\begin{figure}
\begin{center}
\includegraphics[scale=0.64, trim={1.7cm 14.9cm 10.6cm 1.5cm},clip, center]{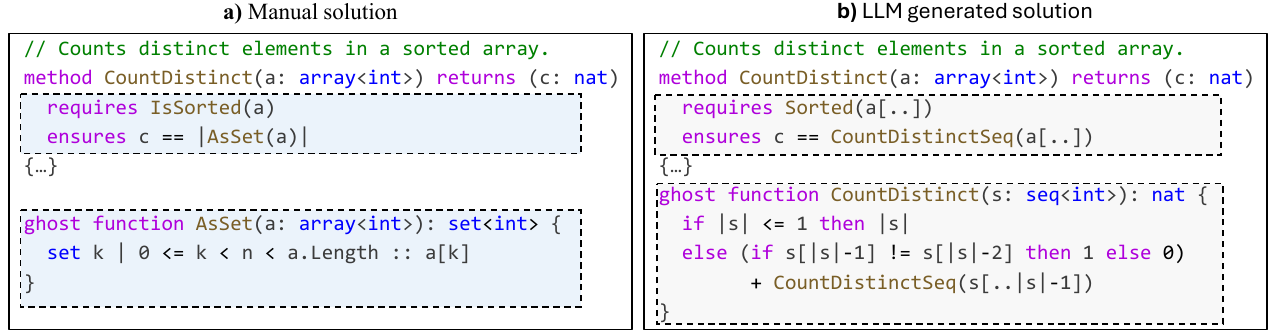}
\end{center}
\vspace{-9pt}
\caption{Example of postcondition generated by Claude Opus 4.5 (on the right) closer to the implementation and less declarative than the manual solution (on the left).}
\label{fig:countdistinct}
\end{figure}

\begin{figure}[H]
\begin{center}
\includegraphics[scale=0.8, trim={1.8cm 6.7cm 16.8cm 0.7cm},clip] {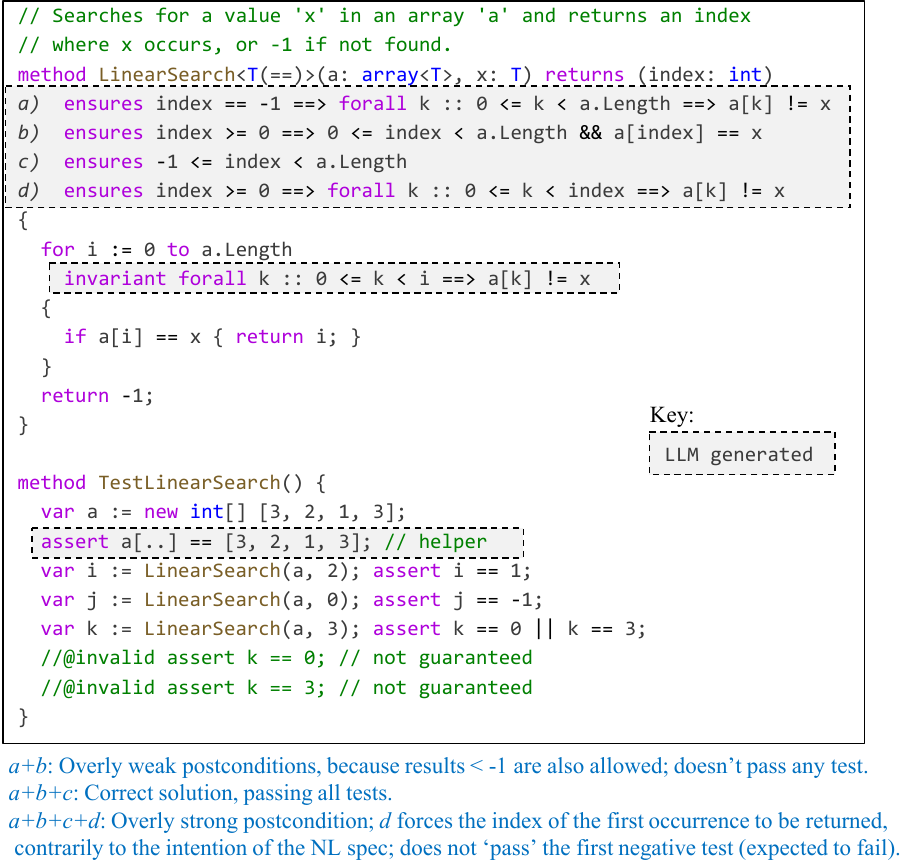}
\end{center}
\vspace{-8pt}
\caption{Example of weak, strong and correct postconditions generated by Claude Opus 4.5 in different attempts. The method body (with the loop invariant also generated by the LLM) is checked successfully in all cases, but the tests reveal the incorrect postconditions. } 
\label{fig:linearsearch}
\end{figure}

For each of the 108 programs successfully solved in the multimodel approach, we selected the minimized version of the first verified solution with fewer LOC. For the 2 unsolved programs, we pick the first version generated by Claude Opus 4.5.

We conclude from the manual analysis that \textbf{the generated pre/postconditions are logically equivalent to expert solutions in 96.4\% of the cases [RQ3]}, with similar complexity in 82.7\% of the cases (textually or structurally identical, or semantically equivalent), higher in 10.0\% (more complex or less declarative,  Fig.~\ref{fig:countdistinct}), and lower (opposite) in 3.6\% of the cases.

The remaining 3.6\% correspond to edge cases with \textbf{overly strong postconditions (1 case) or overly weak preconditions (3 cases)}, consistent with the implementation.
Fig. \ref{fig:linearsearch} illustrates an example of an overly strong postcondition.
An example of an overly weak precondition arises in \texttt{MaxDistEqual(a)}: the specification in natural-language requires a non-empty array, but this constraint is omitted, although empty arrays are safely handled by the implementation.

Such issues can be detected using \emph{negative} test statements that are \emph{expected to fail}, marked \texttt{//@invalid} in Fig.~\ref{fig:linearsearch} and Fig.~\ref{fig:div}. After verifying the LLM-generated solution, we uncomment each negative test in turn and rerun the Dafny verifier; if verification succeeds, we report an error and retry LLM generation using the resulting message. Using this approach on the 4 programs with overly weak preconditions or overly strong postconditions, Claude Opus~4.5 produced correct solutions for all of them.

To further assess test effectiveness, we manually analysed 165 failed attempts generated by Claude Opus~4.5 under repair prompting, excluding syntax errors and verifier timeouts. In 7 failed attempts (see example in Fig.~\ref{fig:linearsearch}), the method body verified but tests failed due to weak or missing postconditions, demonstrating the value of test assertions for detecting specification issues.

Taken together, these results show that \textbf{the original test suites reliably exposed incorrect or incomplete specifications, confirming their adequacy as test oracles. 
In a small number of edge cases involving weak preconditions or strong postconditions, extending the test suites with statements \emph{expected to fail} further improves the detection and resolution of specification issues. [RQ2]}

\newpage

\subsection{Answers to Research Questions}
\label{sec:answers}

\noindent\textbf{RQ1. How effectively can current LLMs generate specification and verification annotations (pre/postconditions, loop invariants, and proof helpers) to successfully verify Dafny programs of moderate complexity?}\\
Our experiments show that current LLMs are highly capable of producing annotations for verifying  Dafny programs of moderate complexity. In the multimodel approach combining Claude Opus 4.5 and GPT-5.2 with repair prompting, 98.2\% of the TESTDAFNY110 programs were solved within at most 8 attempts and just 2 on average. 
Only two complex problems were not successfully solved: \texttt{FastModularExponentiation} and \texttt{PrimeFactorization}, requiring 4 and 10 lemmas in the manual solution, respectively.

\medskip
\noindent\textbf{RQ2. How effective are test methods with test assertions at detecting incorrect or incomplete specifications?}\\
The test assertions in the dataset proved highly effective as specification oracles. 
Weak or incomplete postconditions consistently caused test failures, including cases in which method bodies verified successfully. 
In a small number of edge cases involving weak preconditions or strong postconditions, extending the test suites with statements expected to fail further improved the detection and resolution of specification issues.
Taken together, the test suites exposed all specification defects, demonstrating that test-driven validation is a practical and robust mechanism for specification checking when supported by adequate test suites. 

\medskip
\noindent\textbf{RQ3. How does the quality of LLM-generated solutions compare to expert solutions?}\\
Regarding solution \textit{complexity}, in the multimodel approach with repair prompting, the LLM-generated solutions (for 108 out of 110 programs) contained 63.2\% more LOC than their manual counterparts; by applying a subsequent minimization procedure, this overhead could be reduced to just 10.6\%. 
Regarding solution \textit{correctness}, the generated pre/postconditions were logically equivalent to the expert-written specifications in 96.4\% of the cases; the remaining 3.6\% corresponded to borderline situations of overly weak preconditions or overly strong postconditions, which could be addressed by adding test statements expected to fail.

\medskip
\noindent\textbf{RQ4. What are the main challenges and limitations of LLM-guided generation of specification and verification annotations?}\\
A logistic regression analysis (Section~\ref{sec:factors}) reveals that proof-helper annotations contribute disproportionately to problem difficulty compared to standard specification elements, reflecting the greater challenge these elements pose for current models. In repair iterations, LLMs tend to generate overly complex helpers, a problem that can be significantly mitigated by applying a subsequent minimization step, mimicking human practices. Because of the possibility of cheating (e.g., inserting \texttt{assume} statements, disabling termination checking with \texttt{decreases *}, or simplifying implementation), generated solutions should be manually inspected.

\subsection{Threats to Validity}
\label{sec:limitations}

Although our dataset contains programs of varying complexity, practical problems may be more challenging \cite{faria2023case}, and the performance results may not generalise. 
In the future, we aim to investigate how LLMs can generate specification and verification annotations for larger or more complex programs, by breaking the problem into smaller tasks (divide and conquer approach).

We report results obtained using only very large commercial language models from OpenAI, Anthropic, and DeepSeek via their APIs; therefore, the findings may not generalize to smaller or more cost-efficient models. Future work will explore open-source, locally deployed, and smaller models to study cost--performance trade-offs and identify faster and more affordable solutions.

A potential threat to validity in LLM-based evaluations is unintended exposure to benchmark problems during training. To assess possible data contamination, we evaluated GPT-4, whose knowledge cutoff predates the source repositories used in our TESTDAFNY110 dataset. GPT-4 achieved a pass@5 score only 7.3 percentage points lower than OpenAI’s newest model (GPT-5.2) in non-reasoning mode (29.1\% vs.\ 36.4\%), a gap comparable to contamination-free benchmarks such as LiveBench~\cite{livebench}, suggesting low contamination risk. 
A manual comparison between generated specifications and manual solutions (many previously available in public repositories) revealed no evidence of copying; instead, models appear to learn and apply general specification patterns. 
We further assessed potential contamination by incorporating subdataset membership into the success prediction model (Section~\ref{sec:factors}). After controlling for structural complexity, subdataset origin showed no statistically significant effect (likelihood-ratio test, $\chi^2(2) = 0.79$, $p = 0.67$), providing converging evidence that our results primarily reflect genuine model capabilities and problem complexity rather than training data exposure.

\section{IDE Integration and Usability Study}
\label{sec:vscode}

\subsection{Dafny AI Assistant}
\begin{figure}[H]
\centering
\includegraphics[scale=0.6, trim={1.6cm 4.9cm 11.8cm 1.1cm},clip]{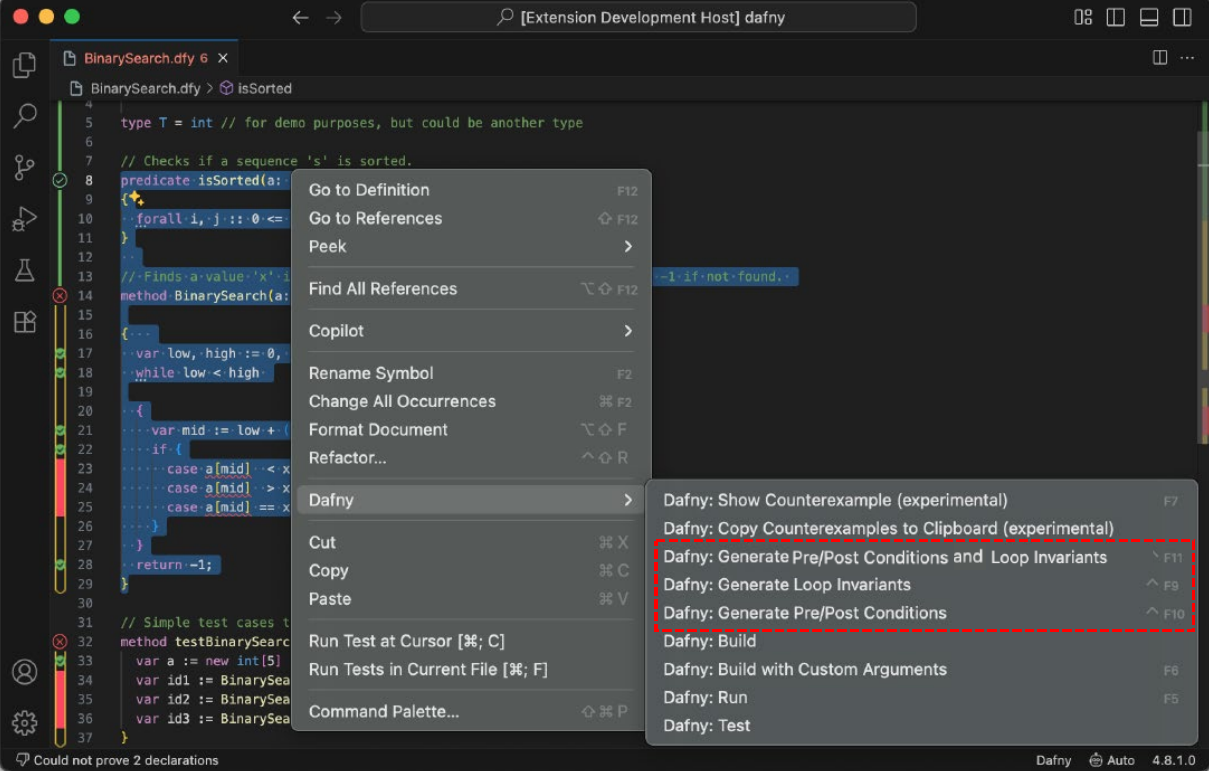}
\vspace{-8pt}
\caption{Options added by our Dafny AI Assistant to the official VSCode Dafny extension.}
\label{fig:vscode}
\end{figure}

We developed a tool extension, named Dafny AI Assistant, publicly available on GitHub\footnote{\url{https://github.com/emantrigo/dafny-plugin}}, that extends the official Dafny extension for Visual Studio Code\footnote{\url{https://github.com/dafny-lang/dafny}} (VSCode) with options for automatic pre/postcondition and/or loop invariant generation directly inside the IDE (see Fig.~\ref{fig:vscode}).

The tool supports LLMs from OpenAI, Anthropic, DeepSeek, and xAI through a unified interface. API keys are supplied securely via Firebase Remote Config. It implements provider prioritization and transparent failover: if a generation attempt fails due to syntax errors,  verification failure, rate limiting, or timeout, it automatically retries with an alternative provider.

Users can invoke generation for a selected fragment (typically a method or a whole file). The tool prompts the configured LLMs to generate or repair annotations, including in the prompt the diagnostic information produced by the verifier. If the result verifies successfully (within a configurable retry limit), the code is updated in the editor. Otherwise, the tool presents to the user the version that verified the most proof obligations and produced the fewest errors, who should then determine what needs to be fixed in the implementation, the annotations, or the test methods.

\subsection{Usability Study}

We conducted a controlled usability study with 7 participants, each performing 6 formal verification tasks of varying complexity (for a total of 42 tasks), both manually and with tool assistance. We measured success rate, completion time, and perceived usability.

All participants were master's students enrolled in a course on Formal Methods for Critical Systems, with six weeks of prior experience with Dafny.

The tasks covered 3 complexity categories, with 2 problems per category:
\begin{itemize}\itemsep2pt \parskip2pt \parsep2pt
%[noitemsep, topsep=3pt]
\item Numerical algorithms (A): Catalan numbers (A1), prime checking (A2);
\item Array queries (B): find minimum (B1), heap property checking (B2);
\item Array mutations (C): selection sort (C1), array reversal (C2).
\end{itemize}

Each task required participants to write missing preconditions, postconditions, loop invariants, and any auxiliary constructs needed to obtain successful Dafny verification. The provided code included test methods with assertions that would only pass with correct formal specifications.

A crossover design controlled for individual skill and task bias: group~1 (4 participants) completed A1, B1, C1 with the tool and A2, B2, C2 manually; group~2 (3 participants) followed the reverse order. Each exercise had a 15-minute time limit, with a 1-hour overall limit, simulating realistic development pressure while allowing meaningful interaction with the tool.

At the end of the session, participants completed a questionnaire including the System Usability Scale (SUS), along with background and qualitative feedback questions, showing clear performance gains when using the tool:
\begin{itemize}\itemsep2pt \parskip2pt \parsep2pt
    \item Success rate: 85.7\% with the tool vs.\ 42.1\% manually;
    \item Average completion time: 6.2~min with the tool vs.\ 11.1~min manually;
    \item SUS score: 74.6/100, indicating good usability.
\end{itemize}

%In the qualitative feedback, participants reported that automatic loop invariant generation was particularly helpful for iterative code, although occasional over-specification and solver timeouts still occurred. The ability to switch providers and use fallback generation was also strongly appreciated.

In summary, the tool substantially improved success rates and completion times, and was perceived as usable and helpful by participants. Although the study was small in scale, the results suggest that AI-assisted specification generation can effectively support practical formal verification tasks.
%, even for users with limited experience.

\section{Related Work}
\label{sec:related}

\subsection{Loop Invariant Synthesis}

Automatically inferring loop invariants is, in general, undecidable and remains a long-standing challenge. Traditional heuristic-guided search approaches struggle with scalability and generalizability \cite{galeotti2015inferring, sharma2016invariant}. More recent work leverages machine learing (ML) and LLMs to address these limitations \cite{si2018learning, janssen2024can, kamath2023finding, pei2023can, wu2024llm}, primarily focusing on C verification benchmarks.

In \cite{si2018learning}, the authors introduce CODE2INV, a neural framework based on reinforcement learning. Evaluated on 133 C programs with preconditions, a loop, and a postcondition, CODE2INV solved 80\% of the problems, outperforming stochastic search (55\%), heuristic search (58\%), and decision tree learning (75\%).

In \cite{janssen2024can}, GPT-3.5 was prompted to regenerate human-written ACSL loop invariants for 106 C programs. With up to five attempts per program, valid invariants were produced for 75 tasks, enabling successful verification of 22 programs with Frama-C~\cite{cuoq2012frama} and 36 with CPAchecker~\cite{beyer2011cpachecker}.

A more extensive study by \cite{kamath2023finding} uses GPT-4 with refined prompting on 469 integer-arithmetic C programs. Correct invariants were found for 50.5\% of the programs using a basic prompt after 15 attempts, increasing to 84.9\% when incorporating invariant patterns, candidate reuse, and verifier feedback. Their approach complements symbolic methods~\cite{heizmann2018ultimate}, with only partial overlap in solved problems. Similarly, \cite{wu2024llm} reports a 97.8\% success rate on 316 C programs by combining candidates from failed attempts using BMC-guided selection.

Compared with these approaches, our work targets a broader class of problems, including integer arithmetic, mutable arrays, reals, strings, and sequences, in a higher-level language and with direct IDE integration. Although overall success rates are comparable, the datasets are not directly comparable.

In \cite{pei2023can}, the authors study invariant prediction for Java programs and show that fine-tuned models achieve quality comparable to Daikon~\cite{ernst2007daikon}. While their loop-body invariants differ from ours (they need not to imply a postcondition), their fine-tuning strategy may be adaptable to our setting.

Recently, \cite{bharti2025loop} reports 100\% success on the CODE2INV benchmark by combining reasoning LLMs with SMT solvers. Likewise, we generate correct loop invariants for all TESTDAFNY110 programs, which cover a broader problem class, including arrays and nested loops.

\subsection{Specification Synthesis and Validation}

LLMs are increasingly used to generate specifications for verification-aware languages and frameworks such as ACSL/Frama-C, Dafny, and JML/OpenJML. Some focus solely on pre/postconditions \cite{lahiri2024evaluating, zhu2025formalizing, richter2025beyond}, while others also synthesize loop invariants \cite{ma2025specgen, wen2024enchanting} or even full implementations~\cite{misu2024towards}.

Misu et al.~\cite{misu2024towards} show that GPT-4 can generate verified Dafny programs (specifications and implementations) for 58\% of an MBPP-derived benchmark (MBPP-DFY), based on natural-language task descriptions and tests. In contrast, our approach starts from a conventional program with test code and focuses on synthesizing the annotations required for verification.

Richter et al.~\cite{richter2025beyond} propose \emph{NL2Contract} to translate natural-language docstrings into contracts expressed as executable Python assertions. Contract quality is evaluated for soundness using CrossHair and approximated completeness via mutation testing on HumanEval+ and Python-by-Contract, with a small manual study confirming alignment with intended behavior.

AutoSpec~\cite{wen2024enchanting} is, like our work, a combined approach that synthesizes pre/postconditions and loop invariants using LLMs and verifier feedback. It relies on ACSL assertions encoding general correctness properties, making it property-driven rather than test-driven. Using an iterative feedback loop, AutoSpec verifies 79\% of 251 C programs within five attempts, but depends on users providing correct and complete properties, making it vulnerable to under-specification. Although the benchmarks differ, our results suggest that combining LLMs with verifiable test oracles is both effective and closer to typical developer workflows.

SpecGen~\cite{ma2025specgen} is another combined approach targeting Java, synthesizing JML specifications checked with OpenJML. It iteratively refines specifications using verifier feedback and a mutation-based repair phase. On a benchmark of 385 Java programs, SpecGen verifies 72.5\% of the programs, outperforming both LLM-based and non-LLM baselines. However, unlike property- or test-driven methods, semantic adequacy is assessed manually.
%, with high agreement reported for 15 inspected solutions.

We next describe two approaches inspired by mutation-testing techniques to validate Dafny specifications, either manually provided or LLM-generated.

MutDafny~\cite{amaral2026mutdafny} introduces mutants into method bodies and checks whether existing method postconditions are able to detect them; mutants that still verify indicate potential specification weaknesses or underspecification. The approach was evaluated on 794 Dafny programs, with a manual analysis of 284 surviving mutants from 24 randomly selected programs, identifying five programs with weak postconditions. 
Notably, four of these programs are also part of our subdataset~A (task\_id\_2, 126, 161, and 249), for which (likewise for all 108 solved problems) our LLM-generated specifications do not exhibit underspecification issues (Sec.~\ref{sec:specquality}).

An advantage of MutDafny is that it does not require test cases to validate Dafny specifications. However, it incurs significant manual effort to inspect surviving mutants (with many false positives) and may miss some specification weaknesses (false negatives). In contrast, our approach shifts the manual effort to writing test cases, which provide strong assurance guarantees and align with standard development workflows. The two approaches are therefore complementary.

\begin{figure}[H]
\begin{center}
\includegraphics[scale=0.45, trim={1.6cm 10.7cm 11cm 1.3cm},clip, center]{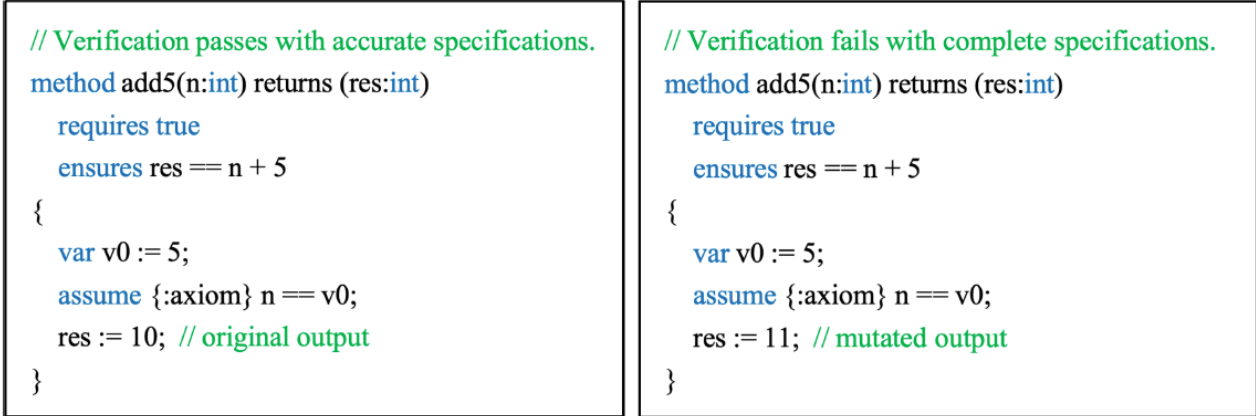}
\end{center}
\vspace{-16pt}
\caption{Mutation testing approach of \cite{zhu2025formalizing}.}
\label{fig:mutationtesting}
\end{figure}

In another work, Lu et al.~\cite{zhu2025formalizing} propose a mutation-testing approach, building on \cite{lahiri2024evaluating}, to validate LLM-generated pre/postconditions for Dafny methods. Given generated preconditions $P$ and postconditions $Q$, their method checks accuracy and completeness against valid input--output pairs $(x_i,y_i)$ by generating mutated outputs $(x_i,y'_{i,j})$ and requiring that $P(x_i)\!\implies\!Q(x_i,y_i)$ holds for all valid pairs, while $P(x_i)\!\implies\!Q(x_i,y'_{i,j})$ fails for all mutants. These checks are discharged by the Dafny verifier using encodings illustrated in Fig.~\ref{fig:mutationtesting}.

In contrast, our approach relies on conventional test methods that invoke the target method on concrete inputs and validate outputs via assertions such as \texttt{assert y == $y_1$ || \dots || y == $y_p$}. A specification passes a test if the precondition holds and the postcondition implies the assertion, formally,
$P(x_1) \;\land\; (\forall y.\; Q(x_1,y) \Rightarrow (y = y_1 \lor \cdots \lor y = y_p))$.
When $p=1$, all output mutations necessarily fail, eliminating the need for mutation testing. When $p>1$, mutation testing may produce false rejections by mutating one valid output into another. In our setting, if the test assertion is verified, no additional outputs are possible, again avoiding mutations. Overly strong postconditions can still be detected using test assertions expected to fail.

Compared to mutation testing, our approach avoids the need to design adequate mutants, supports more flexible test code, and aligns more closely with standard development practices. 
For instance, in MBPP task~426 (filtering odd numbers), their approach accepts a weak specification that omits order preservation, whereas our test-driven approach correctly rejects it.

They evaluated their method on 123 MBPP-DFY tasks using GPT-4 and DeepSeek-R1. On 64 tasks overlapping with TESTDAFNY110, they obtained pass@5 = 87.5\% for pre/postcondition generation, whilst our multimodel approach achieves repair@6 = 100\% for combined generation of pre/postconditions, loop invariants, and proof helpers. 
Beyond using newer models and iterative repair, a key factor in our higher success rate is allowing the LLM to introduce auxiliary helper functions and predicates when needed, which \cite{zhu2025formalizing} explicitly avoid due to misuse concerns.

Overall, our approach differs from prior work by using test methods with statically checked assertions as specification oracles: a candidate specification is accepted only if it verifies and makes all test assertions provable (and negative tests non-provable). While this places higher demands on test quality and proof helpers, it provides stronger guarantees than other approaches.

\subsection{Proof Helper Synthesis}

Recent work has investigated the use of LLMs to reduce the manual effort
required to generate proof helpers, such as auxiliary assertions
or lemmas, which assist the verifier in discharging proof obligations
that cannot be handled automatically, beyond pre/postconditions, and loop invariants.

Early work by Silva et al.~\cite{silva2024leveraging} explored the use of
LLMs to assist Dafny developers by suggesting missing lemmas and proof
steps that Dafny cannot discover automatically. Experiments with
GPT-4 Turbo demonstrated the ability to infer relevant lemmas, but the
model failed to complete full proofs.

Laurel~\cite{mugnier2025laurel} introduces a two-phase approach for
fixing Dafny lemmas by synthesizing assertions guided by verifier
messages and codebase-level similarity. Evaluated on the DafnyGym
benchmark of 143 assertion synthesis tasks extracted from real-world
codebases, Laurel generated 56.6\% of the required assertions within 10
independent attempts per task with GPT-4o, using verifier feedback to localize
insertion points and retrieve similar examples.

Silva et al.~\cite{silva2025inferring} propose DAISY, a pipeline for
automatically inferring missing helper assertions in Dafny programs.
Using GPT-4.1, it verified 63.2\% of programs with a single missing
assertion and 31.7\% with multiple missing assertions on 506 programs
derived from DafnyBench, outperforming Laurel on the same benchmark.
Like Laurel, DAISY follows a two-stage process of fault localization and
assertion inference, generating 10 candidate assertions per location.

Unlike our approach, both Laurel and DAISY rely on a fixed set of
independent candidate generations and do not apply an iterative repair
strategy. Moreover, they focus exclusively on assertion discovery,
without addressing other forms of proof helpers or the synthesis of full
specifications.

Poesia et al.~\cite{poesia2024dafny} propose dafny-annotator, an
LLM-guided tool that automatically adds logical annotations to Dafny
methods until verification succeeds. The approach assumes correct
preconditions and postconditions and focuses on intra-method
annotations. Fine-tuning on a combination of DafnyBench and the
synthetic DafnySynth dataset improves annotation recovery from
15.7\% to 50.6\% with LLaMa 3.1 8B.

In contrast, our approach treats proof helper generation as part of an
\emph{end-to-end iterative repair process} that jointly synthesizes
preconditions, postconditions, invariants, and proof helpers using test
assertions as semantic oracles. Our experimental results demonstrate
the effectiveness of this strategy for proof helper generation.

\section{Conclusions and Future Work}
\label{sec:conclusions}

Our experiments demonstrate that LLMs, particularly through a multimodel approach with repair prompting and test assertions as specification oracles, can achieve high accuracy in generating specification and verification annotations for successful program verification in Dafny, with an 98.2\% success rate within at most 8 attempts in the TESTDAFNY110 dataset. 
%Our VS Code extension (Dafny AI Assistant) integrates such capabilities directly into the IDE.

In contrast to previous approaches, our method introduces three key advances: (i) the use of statically checked test assertions as semantic oracles, which automatically expose incorrect or incomplete specifications and closely align with developer workflows; (ii) a multimodel LLM approach with a \textit{guess-check-repair-minimize} workflow that improves success rate and output quality; and (iii) IDE integration via the Dafny AI Assistant. Together, these elements enable a more robust, practical, and comprehensive solution to automatic pre/postcondition, loop invariant and proof helper generation.

Integrating LLMs into formal verification workflows has the potential to reduce manual effort and promote wider adoption of formal methods.

For future work, we plan to:
(1) extend our pattern catalogue of common problems and solutions for annotation generation;
%(2) improve minimization efficiency through tighter, incremental interaction with the Dafny language server;
(2) explore LLM-based annotation generation for more complex programs using divide-and-conquer task decomposition; and
(3) evaluate open-source, locally deployed, and smaller language models to analyze cost–performance trade-offs.

\section*{Acknowledgments}
J. Pascoal Faria was financed by National Funds through the FCT - Fundação para a Ciência e a Tecnologia, I.P. (Portuguese Foundation for Science and Technology) within the project VeriFixer, with reference 2023.15557.PEX (DOI: 10.54499/2023.15557.PEX).

\section*{Declaration of generative AI and AI-assisted technologies in the manuscript preparation process}

During the preparation of this work the authors used ChatGPT and GitHub Copilot to simplify some phrases and assist in Python script creation. After using these services, the authors reviewed and edited the content as needed and take full responsibility for the content of the published article.
%vspace{4pt}

\appendix
\section{System Prompts}
\label{sec:system-prompts}

\subsection{Direct Prompt}
\label{sec:base_prompt}

\lstset{
  basicstyle=\ttfamily\small,
  frame=single,
  breaklines=true,
  breakindent=0pt
%  breakatwhitespace=true,
%  columns=fullflexible
}
                   
%[caption={System prompt used for generation}, label={lst:system-prompt}]
\begin{lstlisting}[basicstyle=\ttfamily\footnotesize, frame=single, breaklines=true, breakindent=0pt]
You are an expert in the Dafny programming language and formal verification.  
Your task is to insert any missing pre-conditions ('requires' clauses), post-conditions ('ensures' clauses), loop invariants ('invariant' clauses) and any auxiliary ghost predicates, ghost functions, and proof helpers (assertions and lemmas) needed for successful verification of the provided Dafny code, following these specific instructions.

Task Requirements:
- The Dafny code will be enclosed between the tags BEGIN DAFNY and END DAFNY.
- Do not change the original Dafny code! Your task is just to insert 'requires', 'ensures', and 'invariant' clauses (plus auxiliary ghost predicates, ghost functions, and proof helpers, if needed).
- Do not provide any explanations or comments in your output; only output the modified code between the tags BEGIN DAFNY and END DAFNY.

Follow Dafny syntax rules:
- Method preconditions and postconditions must be placed immediately after the method header and before the method body, as in:
   method Abs(x: int) returns (y: int)
     requires true
     ensures y >= 0
     ensures x == y || x == -y
   {
     if x < 0 { return -x; } else { return x; }
   }
- Loop invariants must be placed immediately after the loop header and before the loop body, as in:
   while i < n
     invariant 0 <= i <= n
   { ...
     i := i + 1;
   }
- Use '==>' for logical implication.
- Use |s| for sequence length and a.Length for array length.
- Do not use dot operations on collections like s.Map, s.Contains, s.Min, etc.
- Do not use aggregate functions on collections, like max, min, and sum, unless they are defined in the code.   
- Do not invoke methods inside pre/post-conditions (only functions and predicates can be invoked).
- Use "if then else " for ternary expressions instead of " ? : ".
- The syntax 'function' (specification) 'by method' (implementation) is valid in Dafny.
    
Guidelines for writing pre/post-conditions (requires and ensures clauses):
- Do not add pre/post-conditions to test methods or the Main method.
- Do not add pre/post-conditions for methods defined with 'by method' (as they are inherited from the function/predicate definition).
- Do not add redundant conditions like 'requires true' or 'ensures true'.
- Do not add null checks like 'requires x != null', because by default reference types in Dafny do not accept null.
- Write pre/post-conditions that capture all relevant constraints.
- For requires clauses, specify necessary conditions for the method to work correctly.
- For ensures clauses, thoroughly establish the ranges and constraints that the returned values and output parameters must satisfy.
- Where possible, use quantified expressions (easier to verify) instead of recursive definitions.
- Always provide explicit lower bounds in quantifiers like forall k :: 0 <= k <= n, instead of forall k :: k < n.
- When a method modifies an array ('modifies' clause), consider what parts of the array change and what parts remain unchanged.
- When a method updates object fields, the ensures clause should describe the relationship between pre-state (old) and post-state values.
- If needed for writing thorough pre/post-conditions, you can create auxiliary ghost functions and predicates (namely for recursive definitions).
- In case you need to create a recursive function, make sure that the direction of recursion matches the implementation being verified.
  Example of preferred form if implementation iterates from lower to higher indices: ghost function Sum(s: seq<int>): int {if |s| == 0 then 0 else s[|s|-1] + Sum(s[..|s|-1])}    
  Example of form to avoid: ghost function Sum(s: seq<int>): int {if |s| == 0 then 0 else s[0] + Sum(s[1..])}    
- If there are test assertions on lists (sequences, arrays, strings) and the postconditions under test involve recursive functions, you can set the fuel attribute to the list length as in {:fuel n} to facilitate verification.
  
Guidelines for writing loop invariants:
- Try to create loop invariants with a structure similar to the method post-conditions ('ensures' clause), reusing auxiliary functions or predicates mentioned in those clauses, to be incrementally enforced as the loop progresses.
- Do not reference uninitialized variables or output parameters in the invariants.
- You must first understand the role of each variable in the algorithm, using any comments provided, to properly construct meaningful invariants.
- Create separate invariants for each variable manipulated within the loop, ensuring that each is well-defined.
- Do not include redundant or overly generic invariants.
- Where applicable, prefer sequence operations over quantifiers on arrays.
- Always provide explicit lower bounds in quantifiers like forall k :: 0 <= k <= n, instead of forall k :: k < n.
- When a loop in a method modifies an array ('modifies' clause), a loop invariant should exist for each segment unchanged up to the current iteration, using old().
- 'for' loops do not need a 'decreases' clause nor a loop invariant for the loop index bounds, as they are automatic.
- When 'boolean' variables are manipulated in a loop, the loop invariants should describe the conditions upon which they may be true and false (covering both cases).
- In 'for' loops, the upper bound is exclusive.
- In the case of descending for loops ('downto'), the loop iterator is implicitly decremented at the beginning of the loop body (not at the end).
- If needed, you can also add 'decreases' clauses to help prove loop termination (in most cases, they are inferred by Dafny).

Failure to follow these instructions strictly will result in incorrect output.
\end{lstlisting}

\subsection{Repair Prompt}
\label{sec:repair_prompt}
\lstset{
  basicstyle=\ttfamily\small,
  frame=single,
  breaklines=true,
  breakindent=0pt
%  breakatwhitespace=true,
%  columns=fullflexible
}
                   
\begin{lstlisting}[basicstyle=\ttfamily\footnotesize, frame=single, breaklines=true, breakindent=0pt]
Fix the preconditions, postconditions, and loop invariants (and any other needed annotations, like ghost functions, ghost predicates, assertions and lemmas) in the following program in Dafny so that it can be verified successfully without any errors.
The program contains one or more methods under verification, one or more test methods with test assertions, and possible auxiliary declarations (functions, predicates, etc.).

Task Requirements:
- The Dafny code will be enclosed between the tags BEGIN DAFNY and END DAFNY.
- Current verification errors will be enclosed between the tags BEGIN VERIFICATION ERRORS and END VERIFICATION ERRORS.  
- Output the modified code between the tags BEGIN DAFNY and END DAFNY.
- Do not remove test cases (method calls and output assertion checking) in the test methods (but you can add/fix proof helpers if needed)!
- Do not change the Dafny program under verification (namely, the algorithms in method bodies); your role is just to fix the annotations needed for successful verification!
- Do not use 'assume' (unproved) statements!
- Do not use 'decreases *' clauses to circumvent verification of termination!
- Follow these steps:
  1) Understand the program and the specification and verification annotations;
  2) Understand the errors; 
  3) Determine the root causes of the errors and possible fixes (using hints below when applicable).

Hints for fixing verification errors: 

1) Test assertions serve as oracles for postconditions, which in turn serve as oracles for loop invariants:
  - If test assertions verify successfully, the postconditions are likely correct;
  - If test assertions fail to verify, the postconditions may be too weak or incorrect, or proof helpers are needed to check the test assertions (e.g., additional assertions or lemmas);
  - If postconditions verify successfully, the loop invariants are likely correct;
  - If postconditions fail to verify, the loop invariants may be insufficient or need proof helpers
  - For nested loops, outer loop invariants serve as oracles for inner loop invariants.

2) Ordering-preserving postconditions are often required for tests to pass; they are best defined using an auxiliary sequence comprehension function, such as (with filtering function 'f' and mapping function 'g'):
  ghost function seqc<T,U>(s: seq<T>, f: T -> bool, g: T -> U): seq<U> 
  {
    if s == [] then []
    else if f(s[|s|-1]) then seqc(s[..|s|-1], f, g) + [g(s[|s|-1])]
    else seqc(s[..|s|-1], f, g)
  }

3) Sometimes the solution is correct, but additional proof helpers are required by the Dafny verifier, such as:
a) Provide assertions to help Dafny prove properties involving lists (sequences, arrays or strings), such as:
    assert a[..i+1] == a[..i] + [a[i]];  // appending
    assert a[..i] == a[..k] + a[k..i]; // concatenation
    assert a[..] == a[..a.Length]; // length
b) Provide (non-recursive) postconditions in recursive function definitions, as in:
    ghost function min(s: seq<int>): int
      requires |s| > 0
      ensures min(s) in s && forall k :: 0 <= k < |s| ==> s[k] >= min(s)
    { if |s| == 1 then s[0] else if s[|s|-1] < min(s[..|s|-1]) then s[|s|-1] else min(s[..|s|-1])}
c) If test assertions fail on lists of length N (sequences, arrays or strings) and the postconditions of the method under test involve recursive functions, the first thing to do is to set the fuel attribute to {:fuel N} in the recursive function, as in:
    ghost function {:fuel 5} min(s: seq<int>): int (...)
d) If the verifier warns that it could not find a trigger for a quantifier,  instead of adding a {:trigger} attribute, extract the complex part of the body to an auxiliary predicate, as in:
    Replace: exists i ::  ... complex_expression(args) ...  
    With: ghost predicate p(args) { complex_expression(args) }
          exists i:: ... p(args) ...
e) In recursive function definitions, make sure that the order of recursion is the opposite of the order of iteration in the implementation being verified, to facilitate verification.

4) Sometimes additional proof helpers are needed in the test methods, to help prove the test assertions on test outcomes, such as:
a) Provide auxiliary assertions on the content of arrays to help trigger verification, as in:
    var a := new int[] [1, 3, 5];
    assert a[..] == [1, 3, 5]; // helper
    assert a[0] == 1 && a[1] == 3 && a[2] == 5; // alternative helper 
b) Provide auxiliary assertions with concrete examples, counter examples or intermmediate results to convince Dafny of the test outcomes.
c) Only if previous options don't work, provide auxliary lemmas proving the uniqueness of test results.


\end{lstlisting}

\section{Minimization Procedure}
\label{sec:minimization}

\begin{procedurebox}{Dafny program minimization}
\label{proc:minimization}
\footnotesize

\noindent\textbf{Inputs:}
\begin{itemize}[noitemsep, topsep=0pt, leftmargin=*]
  \item $P_{\mathit{orig}}$: an original Dafny program;
  \item $P_{\mathit{ext}}$: a verifying LLM-generated extended program.
\end{itemize}

\noindent\textbf{Output:}
\begin{itemize}[noitemsep, topsep=0pt, leftmargin=*]
  \item $P_{\mathit{min}}$: a verifying minimized version of $P_{\mathit{ext}}$ that preserves all code from $P_{\mathit{orig}}$.
\end{itemize}

\noindent\textbf{Steps:}
\begin{enumerate}[nosep, leftmargin=*]

\item Determine the \textbf{delta} $\Delta = P_{\mathit{ext}} \setminus P_{\mathit{orig}}$ by computing an LCS-based alignment between $P_{\mathit{orig}}$ and $P_{\mathit{ext}}$, ignoring differences in whitespace and attributes.

\item Determine an ordered set $\mathcal{C}$ of \textbf{removal candidate segments} from $\Delta$:
\begin{enumerate}[nosep, topsep=0pt, leftmargin=*]
 \item single- or multi-line statements in method and lemma bodies, including composite blocks (\texttt{if}, \texttt{while}, \texttt{for}, \texttt{forall}, \texttt{calc}) and nested statements;
  \item parts of composite  blocks from (a): \texttt{else} branches of conditionals; \texttt{by \{...\}} blocks in \texttt{assert ... by \{...\}} statements;
  \item groups of sequential single-line statements from (a) (contiguous in $P_{orig}$);
  \item single-line or multi-line specification clauses (\texttt{invariant}, \texttt{decreases}, \ldots) in loop headers (\texttt{while}, \texttt{for})
  \item single-line or multi-line specification clauses (\texttt{requires}, \texttt{ensures}, \ldots) in declaration headers (methods, lemmas, functions, and predicates).
\end{enumerate}

\item Build a \textbf{dependency relation} $\mathcal{D}$, mapping declarations to referenced declarations, and maintain corresponding reference counters.

\item Initialize the current minimized program as $P_{\mathit{min}} \gets P_{\mathit{ext}}$.

\item Perform multiple rounds until no deletion is committed in a round. In each round, traverse candidates in $\mathcal{C}$ \emph{backwards} (from bottom to top) and \emph{outside-in} (from outer to inner segments). For each candidate $c \in \mathcal{C}$:
\begin{enumerate}[nosep, leftmargin=*]

  \item Check whether the removal candidate $c$, belonging to a declaration $d$, satisfies at least one of the following conditions; otherwise, skip it:
  \begin{itemize}[noitemsep, topsep=0pt, leftmargin=*]
    \item $d$ was modified in the previous or current pass by $d$ (true in round~1);
    \item $d$ references a declaration $d'$ whose header (specification) was modified in the last pass (in this or previous round, depending on relative position);
    \item $c$ belongs to the header (specification) of $d$, and $d$ is referenced by a declaration $d''$ modified in the last pass.
  \end{itemize}

  \item Tentatively construct $P_{\mathit{tent}} \gets P_{\mathit{min}} \setminus c$ by deleting segment $c$, possibly applying a safe syntactic replacement.

\item Check whether $P_{\mathit{tent}}$ is successfully checked by the Dafny verifier with a short timeout (10~s), using the \texttt{----filter-symbol} option to speed up verification of deletions inside method or lemma bodies. Cache verification results to avoid repeated work. If verification fails, skip this candidate.

  \item Commit the deletion ($P_{\mathit{min}} \gets P_{\mathit{tent}}$) and update the candidate set $\mathcal{C}$.

  \item Update reference counters in $\mathcal{D}$ and recursively remove from $P_{\mathit{min}}$ any declarations that are no longer referenced.
\end{enumerate}

\end{enumerate}
\end{procedurebox}

\setlength{\bibsep}{2pt}

\bibliographystyle{elsarticle-num}
\bibliography{references}

\end{document}